\documentclass[12pt]{article}

\usepackage{amsmath,amssymb,amsfonts}
\usepackage{graphicx}
\usepackage{booktabs}
\usepackage{hyperref}
\usepackage{geometry}
\newcommand{\HG}{\mathrm{HG}}
\newcommand{\BTF}{\mathrm{BTF}}

\title{First-return statistics in Henyey--Greenstein scattering:\\
Motzkin polynomials and the Cauchy kernel}

\author{C~Zeller$^1$ and R~Cordery$^2$\\[6pt]
$^1$ Claude Zeller Consulting LLC, Tillamook, Oregon 97134, USA\\
$^2$ Department of Physics, Fairfield University, Fairfield, Connecticut 06824, USA\\[4pt]
E-mail: \texttt{czeller@ieee.org}, \texttt{rcordery@fairfield.edu}}

\date{}

\begin{document}
\maketitle

\begin{abstract}
We study first-return statistics for photons undergoing three-dimensional
Henyey--Greenstein scattering in a semi-infinite medium.
In previous work, we showed that one-dimensional first-passage probabilities
expand in Catalan and Motzkin generating functions.
Extending to three dimensions requires a Boundary Truncation Factor (BTF)
that accounts for the restricted angular phase space imposed by the boundary.
Extensive Monte Carlo simulations are used to determine the BTF empirically
as a function of scattering order and anisotropy.
For moderate anisotropy, the BTF is accurately described by a Cauchy kernel
with parameters depending only on the Henyey--Greenstein asymmetry factor.
This closed-form expression reproduces Monte Carlo results to 1--2\% accuracy
over a broad range of scattering orders.
At higher anisotropy, systematic deviations from the Cauchy form are observed
and can be reduced using a one-parameter generalized kernel.
We further extend the framework to oblique incidence by replacing the
normal-incidence return probability with a Legendre-series formula;
the BTF parameters and Motzkin counting machinery are independent of
incidence angle, so only the anchor point of the algorithm changes.
The resulting framework provides a computationally efficient mapping from
three-dimensional anisotropic transport at arbitrary incidence to
one-dimensional combinatorial first-passage theory.
\end{abstract}


\section{Introduction}
\label{sec:intro}

First-passage problems appear throughout statistical mechanics---in polymer
physics, diffusion-limited aggregation, financial mathematics, and queueing
theory~\cite{Redner2001,Rudnick2004,Spitzer1964}. A clean example arises in
radiative transport: photons entering a scattering medium execute a
three-dimensional random walk and will eventually return to the entry boundary. We
consider a semi-infinite medium occupying the half-space $z>0$, with the entry
boundary at $z=0$; photons enter as a collimated pencil beam at incidence
angle $\theta_{\mathrm{inc}}$ (direction cosine
$\mu_{\mathrm{inc}}=\cos\theta_{\mathrm{inc}}$). Physically, this
corresponds to the reflectance of a narrow collimated beam incident on a
scattering half-space---the probability that the beam is scattered back across
the entry surface, resolved by the number of scattering events. The
statistics of this first-passage event encode how stochastic motion interacts
with geometric constraint.

This is a canonical problem of the scalar radiative transfer equation (RTE).
The quantity we compute---the scattering-order-resolved first-return
probability $P(n,g,\mu_{\mathrm{inc}})$---is more fundamental than the
reflectance itself: it serves as a generating function for reflectance at
arbitrary single-scattering albedo~$a\in[0,1]$, since
$R(g,a)=\sum_{n=2}^{\infty}P(n,g)\,a^n$. The entire framework operates at
pure scattering ($a=1$); absorption enters only through this weighted sum,
which amounts to a discrete Laplace transform. Because $P(n,g)$ depends only
on the phase function geometry, a single evaluation of the framework covers
all albedos simultaneously. The result is an
analytical alternative to Monte Carlo for this geometry: one polynomial
evaluation per scattering order, rather than $10^8$ photon trajectories.

The practical significance lies in imaging and inverse problems. Diffuse
optical tomography, tissue characterization, and remote sensing all require
solving the RTE repeatedly---often thousands of times---while fitting
scattering parameters to measured reflectance data. When each forward
evaluation requires a full Monte Carlo simulation, the computational cost
becomes prohibitive. A closed-form forward model that evaluates in
microseconds rather than minutes transforms the feasibility of these inverse
problems. The same applies to industrial applications such as paper and print
characterization~\cite{Modric2009}, where rapid parameter estimation from
reflectance measurements is standard practice.

Throughout, ``first-return'' means first passage back to $z=0$ from $z>0$
traveling in the $-z$ direction---the photon must exit, not graze inward.

In our previous work~\cite{ZellerCordery2020}, we established
that first-return probabilities in one-dimensional isotropic scattering expand
in Catalan numbers, the combinatorial objects counting Dyck paths. The
reflectance of a semi-infinite Kubelka--Munk medium~\cite{KubelkaMunk1931}
admits the generating function representation
\begin{equation}
R_\infty(S,K) = \frac{1}{2}\cdot\frac{S}{S+K}\cdot
C\!\left(\frac{S^2}{4(S+K)^2}\right)
\label{eq:catalan-gf}
\end{equation}
where $C(x)=\sum_{n=0}^{\infty}C_n x^n$ is the Catalan generating
function~\cite{OEIS2024} and $C_n=(2n)!/(n!(n+1)!)$. This is
distribution-free: it depends only on the zigzag structure, not step lengths.

Forward-peaked scattering introduces ``flat'' steps---events with no
$z$-direction change. The Motzkin extension~\cite{Oste2015,Drake2011} handles
this. The one-dimensional first-return probability with forward scattering is
given by the marginal formula:
\begin{equation}
P_{\mathrm{1D}}^{\mathrm{marg}}(n,r) = \sum_{n_p=1}^{\lfloor n/2\rfloor}
\frac{1}{2^{2n_p-1}}\,
\frac{(n-2)!}{(n-2n_p)!\,n_p!\,(n_p-1)!}\;
r^{2n_p-1}(1-r)^{n-2n_p}
\label{eq:P1Dmarg}
\end{equation}
where $n$ is the scattering order ($n\ge 2$), $n_p$ is the peak count (number
of direction reversals), $r$ is the backward-step probability, and $(1-r)$ is
the non-backward probability. The combinatorial coefficient is the Motzkin
triangle number $T(n-2,n_p-1)$.

The central challenge is extending this framework to three-dimensional
anisotropic scattering governed by the Henyey--Greenstein phase
function~\cite{HenyeyGreenstein1941}. Direct embedding of Motzkin structure
via an effective backscattering coefficient $r_b(g)$ fails; as we show below,
the mapping requires a Boundary Truncation Factor (BTF) that accounts for
geometric constraints at the boundary.

We show that the BTF follows a Cauchy kernel in the scattering order~$n$:
\begin{equation}
\BTF(n,g) = \frac{A(g)}{1+\left(\dfrac{n-n_0}{m_x(g)}\right)^2}
\label{eq:BTF-intro}
\end{equation}
with parameters expressible in terms of the anisotropy factor $g$ alone:
\begin{align}
m_x(g) &= \frac{4g}{1-g} \qquad\text{(width)} \label{eq:mx-intro}\\
A(g) &= 1-\frac{g(1+g)}{2} \qquad\text{(amplitude)} \label{eq:A-intro}\\
n_0 &= 2 \qquad\text{(peak location)} \label{eq:n0-intro}
\end{align}
The simple integer coefficients suggest underlying geometric structure;
however, a first-principles derivation of the Cauchy form remains an open
problem (Section~\ref{sec:conclusions}). We state this as a formal conjecture
in Section~\ref{sec:Cauchy}.

The theory applies for $g\lesssim 2/3$ and $n\ge 2$; above $g\approx 2/3$,
deviations grow gradually but can be corrected by a modified Cauchy kernel
with a shape parameter (Section~\ref{sec:highg}).

A key feature of the framework is that the BTF parameters $A(g)$ and $m_x(g)$
are independent of incidence angle. This allows a clean extension to oblique
incidence (Section~\ref{sec:oblique}): only the single-scattering return
probability---the anchor point of the algorithm---changes, while the entire
Motzkin counting machinery carries through unchanged.

\medskip
\noindent\textbf{Practical scope of the validity range.}\quad
The constraint $g<2/3$ complements rather than competes with biological tissue
optics, where $g\approx 0.9$--$0.98$ is typical~\cite{Binzoni2006}.
Table~\ref{tab:materials} summarizes representative anisotropy factors for
various scattering media. Many industrial and environmental applications fall
within the validity range of the present theory; for higher-anisotropy
materials, the Cauchy kernel form remains valid but parameters benefit from
Monte Carlo calibration (Section~\ref{sec:MC}).

\begin{table}[ht]
\centering
\caption{Representative anisotropy factors $g$ for various scattering media.}
\label{tab:materials}
\begin{tabular}{lcc}
\toprule
Material & $g$ & Reference \\
\midrule
\multicolumn{3}{l}{\textit{Within validity range ($g<2/3$)}} \\
Isotropic scatterers & 0 & Definition \\
Paper / print media & 0.4--0.6 & \cite{Modric2009} \\[4pt]
\multicolumn{3}{l}{\textit{Above validity range ($g>2/3$)}} \\
Human dermis/epidermis & 0.7--0.9 & \cite{Jacques2013} \\
Biological tissue (in vivo) & 0.9--0.98 & \cite{Binzoni2006} \\
Intralipid phantoms & $\approx 0.9$ & Standard value \\
\bottomrule
\end{tabular}
\end{table}

\subsection{Relation to previous work}
\label{sec:relation}

Table~\ref{tab:progression} summarizes the progression from~\cite{ZellerCordery2020}.

From our 2020 paper, we carry forward: (i) the Motzkin polynomial framework
(equation~\eqref{eq:P1Dmarg}); (ii) the distribution-free character of
first-passage combinatorics; and (iii) the connection to classical fluctuation
theory~\cite{Feller1971,Andersen1962}. What is new here is: (i)~identification
of the Cauchy kernel form for the BTF through systematic model selection;
(ii)~determination of the parameters $m_x(g)$ and $A(g)$ from Monte Carlo
fitting; (iii)~a modified Cauchy kernel extending validity to high anisotropy;
and (iv)~extension to oblique incidence via a Legendre-series formula.

\begin{table}[ht]
\centering
\caption{Progression of results from~\cite{ZellerCordery2020} through the
present work.}
\label{tab:progression}
\begin{tabular}{lll}
\toprule
Aspect & Zeller \& Cordery (2020) & Present work \\
\midrule
Dimension & 1D & 3D $\to$ 1D via BTF \\
Scattering & Isotropic + forward bias & Henyey--Greenstein \\
Combinatorics & Catalan $\to$ Motzkin & Motzkin + BTF correction \\
BTF & Not needed & Cauchy kernel (empirical) \\
Incidence & Normal only & Normal and oblique \\
Validity range & --- & $g<2/3$ (Cauchy); $g<0.95$ (modified) \\
Status & Derived & Empirical; derivation open \\
\bottomrule
\end{tabular}
\end{table}

\subsection{Comparison with generalized Kubelka--Munk}
\label{sec:gKM}

Our approach differs from previous 3D extensions of Kubelka--Munk. Sandoval
and Kim~\cite{Sandoval2014,Sandoval2017} extended KM through double spherical
harmonics (DP1), obtaining an $8\times 8$ system for forward and backward power
flow. For isotropic scattering in optically thick media, their generalized KM
achieves errors below 15\% when $z_0\ge 10$ (where $z_0$ is the optical
thickness).

However, DP1 encounters difficulties for anisotropic scattering. Its basis
functions contain only first-order azimuthal harmonics and cannot capture
forward-peaked phase functions. At $g=0.8$, Sandoval and Kim found errors
exceeding 80\% in transmitted power; at higher anisotropy, the approximation
gives negative intensities.

The BTF framework inverts the dimensional strategy: instead of enriching 1D
equations with 3D coupling, we use 1D combinatorics and correct for boundary
truncation. Table~\ref{tab:comparison} compares the two approaches.

\medskip
\noindent\textbf{Practitioner guidance.}\quad
For semi-infinite media with $g\lesssim 2/3$, the BTF framework is more accurate
than DP1: 1D polynomial evaluation achieves sub-2\% accuracy versus 15\%+ for
gKM. For finite slabs or strong absorption, use gKM or Monte Carlo. BTF
excels where a fast forward model is needed: iterative parameter fitting in
diffuse optical tomography, reflectance-based tissue characterization, and
any inverse problem requiring thousands of transport evaluations.

\begin{table}[ht]
\centering
\caption{Comparison of dimensional reduction strategies for semi-infinite media.}
\label{tab:comparison}
\begin{tabular}{lll}
\toprule
Property & gKM (Sandoval \& Kim) & BTF (this work) \\
\midrule
Geometry & Finite slab & Semi-infinite \\
Strategy & 1D $\to$ 3D extension & 3D $\to$ 1D reduction \\
Angular basis & DP1 (4 func./hemisphere) & HG sampling + Motzkin \\
Isotropic limit & $<$15\% error for $z_0\ge 10$ & Recovers Catalan structure \\
Anisotropic range & $g\lesssim 0.5$ (qualitative) & $g<2/3$ ($<$2\% deviation) \\
High-$g$ behavior & Unphysical (negative values) & Predictable drift \\
Incidence & Oblique via eigenmode & Normal and oblique \\
Computational form & $8\times 8$ PDE system & 1D polynomial evaluation \\
\bottomrule
\end{tabular}
\end{table}

\subsection{Paper organization}
\label{sec:organization}

Section~\ref{sec:1D} reviews the 1D Motzkin framework~\cite{ZellerCordery2020}.
Section~\ref{sec:BTF} introduces the Boundary Truncation Factor and its Cauchy
kernel form. Section~\ref{sec:MC} describes the Monte Carlo procedure.
Section~\ref{sec:algorithm} presents the computational algorithm for normal
incidence. Section~\ref{sec:oblique} extends the framework to oblique incidence.
Section~\ref{sec:highg} presents the modified Cauchy kernel for
high-anisotropy scattering ($g>2/3$). Section~\ref{sec:conclusions} concludes.

\section{One-dimensional theory: Catalan and Motzkin structures}
\label{sec:1D}

We review results from~\cite{ZellerCordery2020}; see~\cite{Redner2001,Rudnick2004}
for background.

\subsection{Kubelka--Munk reflectance and Catalan numbers}
\label{sec:catalan}

The Kubelka--Munk equations~\cite{KubelkaMunk1931,Myrick2011} describe
one-dimensional radiative transport with isotropic scattering:
\begin{equation}
\begin{pmatrix} dI \\ dJ \end{pmatrix}
= \begin{pmatrix} -(S+K) & S \\ -S & (S+K) \end{pmatrix}
\begin{pmatrix} I \\ J \end{pmatrix} dz
\label{eq:KM}
\end{equation}
where $I$ and $J$ are forward and backward fluxes, $S$ the scattering
coefficient, and $K$ the absorption coefficient. The reflectance of a
semi-infinite slab is
\begin{equation}
R_\infty(S,K) = \frac{S+K}{S}
- \sqrt{\left(\frac{S+K}{S}\right)^2-1}
\label{eq:Rinf}
\end{equation}

The key result of~\cite{ZellerCordery2020} is that this reflectance expands in
Catalan numbers:
\begin{equation}
R_\infty(S,K) = \sum_{n_p=1}^{\infty}
\frac{C_{n_p-1}}{2^{2n_p-1}}
\left(\frac{S}{S+K}\right)^{2n_p-1}
\label{eq:Catalan-expansion}
\end{equation}

Photon trajectories form zigzag random walks; first passage occurs at the
first backward crossing of $z=0$. The probability at the $n_p$-th peak is
$P(n_p)=C_{n_p-1}/2^{2n_p-1}$, connecting to Spitzer's
identity~\cite{Spitzer1964} and Andersen's equivalence
principle~\cite{Andersen1962}.

\subsection{Motzkin extension for forward scattering}
\label{sec:motzkin}

Catalan numbers count Dyck paths---walks restricted to up- and down-steps.
Forward scattering introduces flat steps (events with no $z$-direction
change), requiring the Motzkin extension~\cite{Oste2015,Drake2011,Simon2003}.

\medskip
\noindent\textbf{Definition (Motzkin polynomial).}\quad
The Motzkin polynomial of degree $n$ is
\begin{equation}
M_n(t) = \sum_{k=0}^{\lfloor n/2\rfloor} T(n,k)\cdot t^{n-2k}
\label{eq:Motzkin-poly}
\end{equation}
where $T(n,k)$ are the Motzkin triangle coefficients~\cite{OEIS2024}:
\begin{equation}
T(n,k) = \frac{n!}{(n-2k)!\,k!\,(k+1)!}
\label{eq:Motzkin-triangle}
\end{equation}

The one-dimensional first-return probability with forward scattering is
equation~\eqref{eq:P1Dmarg}, encoding two processes: backward steps
(probability $r$) and non-backward steps (probability $1-r$), with the
Motzkin triangle coefficients counting the combinatorial arrangements.

\section{The Boundary Truncation Factor}
\label{sec:BTF}

The Motzkin framework applies to one-dimensional scattering. To use it for
three-dimensional Henyey--Greenstein transport, we need a mapping---an
effective backscattering coefficient $r_b(g,n)$ that makes the 1D formula
reproduce 3D first-return probabilities. This section introduces the Boundary
Truncation Factor (BTF), the correction that makes this mapping work.

\subsection{Physical origin}
\label{sec:BTF-origin}

In bulk scattering without boundaries, Pfeifer and
Chapman~\cite{Pfeifer2008} proved that the Henyey--Greenstein phase function
is closed under successive scattering: after $n$ scattering events, the
angular distribution of the scattering cosine
$\cos\theta=\hat\Omega_0\cdot\hat\Omega_n$ is itself a Henyey--Greenstein
distribution with parameter $g^n$:
\begin{equation}
p_n(\cos\theta;g) = P_{\HG}(\cos\theta;\,g^n)
\label{eq:closure}
\end{equation}
This closure property---not merely the statement
$\langle\cos\theta\rangle=g^n$---underlies the dimensional reduction from 3D
to 1D.

First-passage problems impose geometric constraints that modify this
relationship. The requirement that photons return to their entry boundary
restricts the accessible angular phase space: trajectories that wander too
far forward are less likely to return. This geometric truncation breaks the
closure property, effectively reducing the asymmetry parameter beyond the
bulk value:
\begin{equation}
g_{\mathrm{eff}}^{(\mathrm{constrained})} = g^n\cdot\BTF(n,g)
\label{eq:geff}
\end{equation}
where $\BTF\le 1$ represents the multiplicative reduction due to boundary
constraints. In the bulk limit (no boundary), $\BTF=1$ and we recover
equation~\eqref{eq:closure}; at a boundary, the incomplete angular integration
yields $\BTF<1$.

Formally, the BTF emerges from boundary-constrained angular integration:
\begin{equation}
\BTF(n,g) = \frac{\int\cdots\int\prod_{i=1}^{n}P_{\HG}(\mu_i;g)
\times[\text{return constraint}]\,d\mu_1\cdots d\mu_n}
{\int\cdots\int\prod_{i=1}^{n}P_{\HG}(\mu_i;g)\,d\mu_1\cdots d\mu_n}
\label{eq:BTF-formal}
\end{equation}
where the constraint ensures the photon crosses below $z=0$ after exactly $n$
scattering events. These nested integrals become analytically intractable
beyond $n=3$.

\subsection{Empirical discovery}
\label{sec:BTF-empirical}

Since direct evaluation of equation~\eqref{eq:BTF-formal} is impractical, we
determined BTF empirically. Monte Carlo simulations provide exact first-return
probabilities for 3D photon transport. The simulations covered
$g\in[0.05,0.95]$ (19 values) and $n\in[2,100]$ (99 values) with $10^8$
trajectories per $g$ value and 10 independent runs---approximately $10^{10}$
photon histories ($10^{12}$ scattering events, $\sim$500~CPU-hours).

For each $(g,n)$ pair, we extracted the BTF as the value needed to make the 1D
formula reproduce the Monte Carlo 3D result:
\begin{equation}
P_{3\mathrm{D}}^{(\mathrm{MC})}(n,g)
= P_{\mathrm{1D}}^{\mathrm{marg}}(n,\,r_b(g,n))
\label{eq:BTF-extraction}
\end{equation}
where $r_b(g,n)$ depends on BTF through the effective anisotropy.

\subsection{The Cauchy kernel}
\label{sec:Cauchy}

Systematic model selection---starting with high-order Pad\'e approximants
(rational polynomial fits) and progressively reducing complexity while
monitoring cross-validation error---revealed that the optimal functional form
is a Cauchy kernel.

\medskip
\noindent\textbf{Model selection details.}\quad
The BTF was initially parameterized as a rational approximant in~$g$ and~$n$
with basis vector $\mathbf{v}(g)=[1,g,g^2,\ldots,g^6]^{\!\top}$, giving a
degree-6 Pad\'e approximant in~$g$. Fitting was performed simultaneously
against all $19\times 99$ Monte Carlo $(g,n)$ pairs using both analysis of
variance (ANOVA) and nonlinear least squares (NLS). Complexity was reduced
progressively by dropping higher-order terms while monitoring cross-validation
error. The minimal form retaining full accuracy uses the reduced basis vector
$\mathbf{v}(g)=[1,g,g^2]^{\!\top}$ with coefficient vectors $\mathbf{P}$,
$\mathbf{Q}$, $\mathbf{R}$:
\begin{equation}
\Psi(g,n) = g^2\cdot
\frac{\mathbf{P}\cdot\mathbf{v}(g)}
     {\mathbf{Q}\cdot\mathbf{v}(g) + \mathbf{R}\cdot\mathbf{v}(g)\cdot(n-2)^2}
\label{eq:Pade-Ansatz}
\end{equation}
Both ANOVA and NLS yielded coefficient estimates consistent with the integer
set $\mathbf{P}=[16,-8,-8]$, $\mathbf{Q}=[0,0,16]$, $\mathbf{R}=[1,-2,1]$;
the integer values fall within one standard deviation of the fitted values
for all nine coefficients under both methods. The quadratic $(n-2)^2$
structure in the denominator---which produces the Cauchy form---emerged from
the Pad\'e fitting rather than being assumed. Substituting these
integer-rounded coefficients, the rational form~\eqref{eq:Pade-Ansatz}
collapses algebraically to the closed-form Cauchy kernel below, with
$A(g)=(1-g)(2+g)/2$ and $m_x(g)=4g/(1-g)$, confirming that the
integer structure is exact, not merely a numerical approximation.

\medskip
The optimal functional form is therefore:
\begin{equation}
\BTF(n,g) = \frac{A(g)}{1+\left(\dfrac{n-2}{m_x(g)}\right)^2}
\label{eq:BTF-Cauchy}
\end{equation}
with parameters:
\begin{align}
A(g) &= 1-\frac{g(1+g)}{2} = \frac{(1-g)(2+g)}{2}
\qquad\text{(amplitude)} \label{eq:A}\\
m_x(g) &= \frac{4g}{1-g} \qquad\text{(width)} \label{eq:mx}
\end{align}
The peak location $n_0=2$ corresponds to the minimum scattering order for
first return.

\medskip
\noindent\textbf{Conjecture (Cauchy BTF).}\quad
\textit{For Henyey--Greenstein scattering with anisotropy factor $g<2/3$, the
Boundary Truncation Factor $\mathrm{BTF}(n,g)$ is numerically
indistinguishable, within Monte Carlo precision, from a Cauchy kernel of the
form~\eqref{eq:BTF-Cauchy} with parameters~\eqref{eq:A}--\eqref{eq:mx}. At
present this is supported by Monte Carlo evidence
(Section~\ref{sec:MC}) but lacks a rigorous derivation.}

\medskip
This parameterized form reproduces Monte Carlo-derived BTF values with mean
deviation $<$2\% and cross-validated $R^2>0.999$ for $g\le 2/3$.

\medskip
\noindent\textbf{Statistical evidence for integer coefficients.}\quad
The 10 independent Monte Carlo runs allow statistical testing of the numerical
coefficients. Fitting across all runs yields $n_0=2.2\pm 0.3$ and width
prefactor $3.9\pm 0.3$; the integers 2 and 4 both lie within one standard
deviation of the fitted values. (The exponent~2 in the denominator emerged
from the Pad\'e approximant model selection, not from parameter fitting.) This
suggests exact integer structure rather than numerical coincidence.

\medskip
\noindent\textbf{Limiting behavior.}\quad
For short paths ($n-2\ll m_x$), $\BTF\approx A(g)$ with minimal boundary
effects. For long paths ($n-2\gg m_x$), $\BTF\to 0$ as boundary truncation
dominates. At $g=0$ (isotropic), $A=1$ and $m_x=0$, so $\BTF=1$ for $n=2$
and $\BTF=0$ for $n>2$---reflecting that isotropic scattering requires
exactly two steps for first return.

\begin{figure}[ht]
\centering
\includegraphics[width=0.75\textwidth]{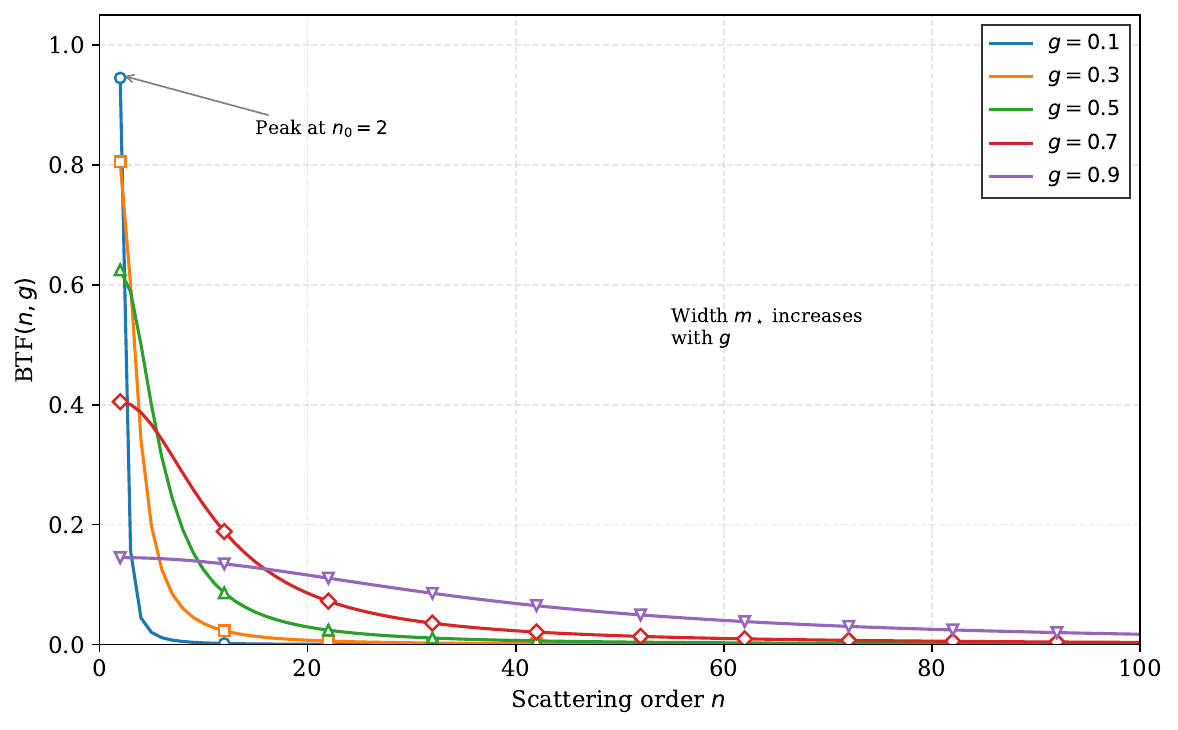}
\caption{The Boundary Truncation Factor versus scattering order $n$ for
$g=0.1,0.3,0.5,0.7,0.9$. Points: extracted from Monte Carlo. Curves: Cauchy
kernel (equation~\protect\eqref{eq:BTF-Cauchy}). The width $m_x(g)$ increases
with $g$; the amplitude $A(g)$ decreases. At $g=0.9$, deviations from Cauchy
become visible.}
\label{fig:BTF}
\end{figure}

\section{Monte Carlo procedure}
\label{sec:MC}

This section details the Monte Carlo simulations from which the BTF was
extracted.

\subsection{Simulation procedure}
\label{sec:MC-procedure}

We follow the standard Monte Carlo procedure for photon
transport~\cite{Jacques2010}. First return occurs when $z$ first becomes
negative~\cite{Sassaroli1998}. Photons are initialized at the origin
($z=0^+$) with direction cosine $\mu_0=1$. At each scattering event, path
length is sampled from $p(\ell)=e^{-\ell}$ (unit mean free path) and
scattering angle from the Henyey--Greenstein phase function via inverse CDF.
The scattering order $n$ is recorded at first occurrence of $z<0$. We
simulated $10^8$ trajectories per $g$ value with 10 independent runs.

\medskip
\noindent\textbf{Operational definition of $n$ and the first-return event.}\quad
The scattering order~$n$ counts every scattering event, \emph{including the
terminal one}. After each scattering event, the cumulative $z$-coordinate is
updated and tested; first return is declared at the first event for which
$z<0$. The terminal scatter is therefore included in~$n$, so the minimum
possible first-return order is $n=2$: one forward scatter followed by one
backward scatter that carries the photon across $z=0$. This definition maps
directly onto the 1D Motzkin path minimum of two steps, with the terminal
downward step corresponding to the first backward crossing. In Monte Carlo
terminology the scattering order equals the path length $m_s$ (see notation
table, Appendix~\ref{app:notation}); both count the same discrete events.

The Henyey--Greenstein phase function is
\begin{equation}
P_{\HG}(\mu;g) = \frac{1}{2}\,
\frac{1-g^2}{(1+g^2-2g\mu)^{3/2}}\,,
\qquad -1\le\mu\le 1
\label{eq:PHG}
\end{equation}
where $g=\langle\mu\rangle\in[0,1)$ is the anisotropy parameter. The
normalization $\int_{-1}^{1}P_{\HG}(\mu;g)\,d\mu=1$ is ensured by the
prefactor~$1/2$.

\subsection{Fit quality}
\label{sec:fit-quality}

The Cauchy kernel captures the Monte Carlo data with mean deviation $<$2\% for
$g<2/3$. Table~\ref{tab:fit} shows the root-mean-square error (RMSE) and
coefficient of determination ($R^2$) at representative $g$ values.

\begin{table}[ht]
\centering
\caption{Fit quality of the Cauchy BTF versus Monte Carlo.}
\label{tab:fit}
\begin{tabular}{cccc}
\toprule
$g$ & RMSE & $R^2$ & Max deviation \\
\midrule
0.10 & $3.2\times 10^{-4}$ & 0.9998 & 1.1\% \\
0.30 & $4.1\times 10^{-4}$ & 0.9997 & 1.4\% \\
0.50 & $4.8\times 10^{-4}$ & 0.9996 & 1.8\% \\
$\frac{2}{3}$ & $5.2\times 10^{-4}$ & 0.9994 & 2.1\% \\
0.80 & $8.3\times 10^{-4}$ & 0.9985 & 5.2\% \\
0.90 & $2.2\times 10^{-3}$ & 0.9941 & 14.8\% \\
\bottomrule
\end{tabular}
\end{table}

\noindent\textbf{Error bounds.}\quad
For $g<2/3$, root-mean-square deviation is below 2\% across all $n$. For
$g>2/3$, errors grow approximately as
$\delta(g)\approx 0.05\cdot(g-2/3)/(1-g)$, reaching $\sim$15\% at $g=0.9$.
For $g>0.8$, the modified kernel (Section~\ref{sec:highg}) is recommended.

\section{Computational algorithm: normal incidence}
\label{sec:algorithm}

We present the complete algorithm for mapping 3D Henyey--Greenstein scattering
to 1D Motzkin combinatorics at normal incidence ($\mu_{\mathrm{inc}}=1$). The
algorithm has four steps: (1)~compute the single-scattering return probability;
(2)~determine the angular threshold; (3)~evaluate the effective backscattering
probability; (4)~apply the Motzkin formula.

\medskip
\noindent\textbf{Step~1 (Single-scattering return probability):}\quad
The two-step return probability anchors the entire framework:
\begin{equation}
p_{r2}(g) = \int_{-1}^{0}\frac{\mu}{\mu-1}\,P_{\HG}(\mu,g)\,d\mu
\label{eq:pr2}
\end{equation}
The geometric factor $\mu/(\mu-1)$ weights the phase function by the
probability of returning to $z=0$ given exit direction~$\mu$.

\medskip
\noindent\textbf{Step~2 (Angular threshold):}\quad
The threshold $\mu_b(g)$ is the direction cosine separating ``effective
backward'' from ``effective forward'' scattering in the 1D projection.
Physically, $\mu_b$ partitions the angular phase space so that the 1D Motzkin
counting matches the 3D return statistics. Solve the self-consistency
condition:
\begin{equation}
p_{r2}(g) = \tfrac{1}{2}\,F(-\mu_b;\,g^2)
\label{eq:mub}
\end{equation}
where $F(\mu;g)$ is the Henyey--Greenstein CDF (cumulative distribution
function):
\begin{equation}
F(\mu;g) = \int_{-1}^{\mu}P_{\HG}(\mu';g)\,d\mu'
= \frac{1-g^2}{2g}\left[
\frac{1}{\sqrt{1+g^2-2g\mu}}-\frac{1}{1+g}\right]
\label{eq:CDF}
\end{equation}
with boundary values $F(-1;g)=0$ and $F(+1;g)=1$.

The parameter $g^2$ appears because after two scattering events the angular
distribution has parameter $g^2$ (equation~\eqref{eq:closure}).

\medskip
\noindent\textbf{Step~3 (Effective backscattering):}\quad
In bulk scattering, the effective anisotropy after $n$ events is $g^n$
(standard convolution). Near a boundary, transverse filtering reduces
this---captured by the BTF. The effective backscattering probability is
\begin{equation}
r_b(g,n) = F\!\left(-\mu_b(g);\,g^{\,2\,+\,\BTF(n,g)\,(n-2)}\right)
\label{eq:rb}
\end{equation}
The exponent interpolates between $2$ at $n=2$ (where $\BTF=A(g)$ is maximal
and the constraint is weakest) and $2$ as $n\to\infty$ (where
$\BTF\to 0$ and boundary truncation dominates). The baseline of $2$
reflects the minimum scattering order for first return. This ansatz is
empirically motivated: it produces the correct limits and matches Monte Carlo
across the parameter space.

\medskip
\noindent\textbf{Step~4 (First-return probability):}\quad
The 3D first-return probability follows from equation~\eqref{eq:P1Dmarg} with
the effective backscattering:
\begin{equation}
P_{3\mathrm{D}}^{(\mathrm{refl})}(g,n)
= P_{\mathrm{1D}}^{\mathrm{marg}}(n,\,r_b(g,n))
= \sum_{n_p=1}^{\lfloor n/2\rfloor}
\frac{1}{2^{2n_p-1}}\,
\frac{(n-2)!}{(n-2n_p)!\,n_p!\,(n_p-1)!}\;
r_b^{2n_p-1}(1-r_b)^{n-2n_p}
\label{eq:P3D}
\end{equation}
This maps 3D Henyey--Greenstein scattering to 1D Motzkin combinatorics.

\medskip
\noindent\textbf{Isotropic limit check.}\quad
At $g=0$: $m_x=0$, $A=1$, so $\BTF=1$ for $n=2$ and $\BTF=0$ for $n>2$.
The exponent in equation~\eqref{eq:rb} becomes $2$ for all $n$. The threshold
$\mu_b(0)=0$ (hemisphere boundary), giving $r_b=1/2$. Substituting into
equation~\eqref{eq:P3D} with $r_b=1/2$ recovers the Catalan first-return
probabilities of~\cite{ZellerCordery2020}, confirming internal consistency.

\section{Extension to oblique incidence}
\label{sec:oblique}

The algorithm of Section~\ref{sec:algorithm} assumes normal incidence
($\mu_{\mathrm{inc}}=\cos\theta_{\mathrm{inc}}=1$). Here we extend the
framework to arbitrary incidence angle. The central result is that only
Step~1---the single-scattering return probability---changes; the BTF
parameters $A(g)$ and $m_x(g)$ are independent of incidence angle, and the
Motzkin counting machinery (Steps~2--4) carries through unchanged.

\subsection{Oblique single-scattering return probability}
\label{sec:oblique-pr2}

At oblique incidence, the photon enters with direction cosine
$\mu_{\mathrm{inc}}\in(0,1]$ rather than $\mu_{\mathrm{inc}}=1$. The
geometric factor governing single-scattering return generalizes to
\begin{equation}
G(\mu_{\mathrm{inc}},\mu_z) = \frac{\mu_z}{\mu_z - \mu_{\mathrm{inc}}}
= \frac{|\mu_z|}{|\mu_z|+\mu_{\mathrm{inc}}}
\label{eq:geom-factor}
\end{equation}
where $\mu_z\in[-1,0]$ is the $z$-component of the exit direction after the
first scattering event. At normal incidence ($\mu_{\mathrm{inc}}=1$), this
reduces to $\mu/(\mu-1)$ as in equation~\eqref{eq:pr2}.

\begin{quote}
\textbf{Important note on the geometric factor.}\quad
The correct form is $G=\mu_z/(\mu_z-\mu_{\mathrm{inc}})$. An earlier version
of this work incorrectly included a spurious factor of $\mu_{\mathrm{inc}}$ in
the numerator:
$G_{\mathrm{wrong}}=\mu_{\mathrm{inc}}\cdot\mu_z/(\mu_z-\mu_{\mathrm{inc}})$.
The error is invisible at normal incidence but introduces a factor of
$\mu_{\mathrm{inc}}$ error at oblique angles.
\end{quote}

The oblique return probability is computed via a Legendre series that
exploits the known expansion of the Henyey--Greenstein phase function:
\begin{equation}
P_{\HG}(\mu;g) = \frac{1}{2}\sum_{l=0}^{\infty}(2l+1)\,g^l\,P_l(\mu)
\label{eq:Legendre-expansion}
\end{equation}
where $P_l$ are Legendre polynomials. The oblique return probability is
\begin{equation}
p_{r2}(g,\mu_{\mathrm{inc}}) = \frac{1}{2}\sum_{l=0}^{L}
(2l+1)\,g^l\,P_l(\mu_{\mathrm{inc}})\cdot I_l(\mu_{\mathrm{inc}})
\label{eq:pr2-oblique}
\end{equation}
where the modal geometric integral is
\begin{equation}
I_l(\mu_{\mathrm{inc}}) = \int_{-1}^{0}
\frac{\mu_z}{\mu_z-\mu_{\mathrm{inc}}}\,P_l(\mu_z)\,d\mu_z
\label{eq:Il}
\end{equation}
The sum is truncated at $L$ sufficiently large for convergence (typically
$L\sim 50$--100).

\subsection{Verification}
\label{sec:oblique-verify}

Three checks confirm the oblique formula:
\begin{enumerate}
\item At $\mu_{\mathrm{inc}}=1$, the mode-sum reduces to the normal-incidence
integral~\eqref{eq:pr2}, since $P_l(1)=1$ for all~$l$.
\item At $g=0$ (isotropic scattering),
$p_{r2}=\frac{1}{4}\cdot\frac{1}{1+\mu_{\mathrm{inc}}}$, which is the
analytical result for isotropic single-scattering return at oblique incidence.
\item $p_{r2}$ increases with decreasing $\mu_{\mathrm{inc}}$ (grazing
incidence makes return easier), consistent with physical intuition.
\end{enumerate}

\subsection{Reference values}
\label{sec:oblique-values}

Table~\ref{tab:oblique-pr2} gives reference values at $g=2/3$ for several
incidence angles.

\begin{table}[ht]
\centering
\caption{Oblique incidence reference values at $g=2/3$.}
\label{tab:oblique-pr2}
\begin{tabular}{cccc}
\toprule
$\theta_{\mathrm{inc}}$ & $\mu_{\mathrm{inc}}$ & $p_{r2}$ & $\mu_b$ \\
\midrule
$0^\circ$ & 1.0000 & 0.025792 & 0.6557 \\
$30^\circ$ & 0.8660 & 0.032594 & 0.5776 \\
$45^\circ$ & 0.7071 & 0.044507 & 0.4512 \\
$60^\circ$ & 0.5000 & 0.071907 & 0.2040 \\
\bottomrule
\end{tabular}
\end{table}

As expected, the return probability increases at oblique incidence (the photon
enters at a shallower angle and is more likely to scatter back out), while the
angular threshold $\mu_b$ decreases (the backward cone opens).

\subsection{Algorithm modification}
\label{sec:oblique-algorithm}

The oblique-incidence algorithm is identical to the normal-incidence algorithm
of Section~\ref{sec:algorithm}, except that Step~1 is replaced:

\medskip
\noindent\textbf{Step~1 (modified):}\quad
Compute $p_{r2}(g,\mu_{\mathrm{inc}})$ using the mode-sum
formula~\eqref{eq:pr2-oblique} instead of the normal-incidence
integral~\eqref{eq:pr2}.

\medskip
\noindent Steps~2--4 proceed identically, using the oblique $p_{r2}$ value.
The angular threshold $\mu_b$ now depends on $\mu_{\mathrm{inc}}$ through
$p_{r2}$, but the BTF parameters $A(g)$ and $m_x(g)$ remain unchanged. The
oblique extension modifies only the anchor point; the Motzkin counting
machinery is independent of incidence geometry.

\section{High-anisotropy extension: modified Cauchy kernel}
\label{sec:highg}

The Cauchy BTF conjecture (Section~\ref{sec:Cauchy}) applies for $g<2/3$.
Biological tissue has $g\approx 0.9$--$0.98$~\cite{Jacques2013,Binzoni2006},
outside this range. Here we develop a one-parameter extension that reduces
errors by roughly half at high~$g$.

\subsection{Systematic deviations at high $g$}
\label{sec:highg-deviations}

Data from $10^{12}$ scattering events show systematic deviations from Cauchy
at high~$g$. For $g>0.85$, Monte Carlo results exceed the Cauchy fit near the
peak and fall below it in the tail---lighter tails than Cauchy.

One possible interpretation: strongly forward-peaked scattering may suppress
the long, wandering trajectories that populate the Cauchy tail. Photons that
scatter many times at high $g$ tend to propagate forward and are less likely to
return (ballistic regime); those that do return must do so in relatively few
steps. This remains conjecture; a rigorous derivation is lacking.

\subsection{Modified Cauchy kernel}
\label{sec:modified-Cauchy}

Generalizing the Cauchy exponent from $1$ to $(1+\alpha)/2$, where $\alpha=1$
recovers the standard form, gives the modified BTF:
\begin{equation}
\BTF_\alpha(n,g) = \frac{A(g)}{\left[1+\left(\dfrac{n-2}{m_x(g)}\right)^2
\right]^{(1+\alpha(g))/2}}
\label{eq:BTF-modified}
\end{equation}
with $A(g)$ and $m_x(g)$ unchanged; $\alpha>1$ gives faster tail decay. This ``generalized Cauchy'' form
appears in robust statistics~\cite{Carrillo2010,Alzaatreh2016}. Note that
$\alpha$ is a shape parameter, not a L\'evy stability index.

\subsection{Fitting the shape parameter}
\label{sec:alpha-fit}

We fit $\alpha$ at each $g$ by least squares against Monte Carlo. The results
follow
\begin{equation}
\alpha(g) = 1 + 0.033\cdot\frac{g-\frac{2}{3}}{1-g}
\label{eq:alpha}
\end{equation}
Table~\ref{tab:alpha} compares fitted and calculated values. The parameter
stays close to unity: $\alpha\approx 0.98$ at low~$g$, rising to
$\alpha\approx 1.2$ at $g=0.95$.

\begin{table}[ht]
\centering
\caption{Fitted $\alpha$ versus formula~\eqref{eq:alpha}.}
\label{tab:alpha}
\begin{tabular}{ccccc}
\toprule
$g$ & $m_x$ & $\alpha$ (fitted) & $\alpha$ (formula) & $(1+\alpha)/2$ \\
\midrule
0.10 & 0.4 & 0.981 & 0.979 & 0.99 \\
0.30 & 1.7 & 0.990 & 0.983 & 0.99 \\
0.50 & 4.0 & 0.990 & 0.989 & 0.99 \\
$\frac{2}{3}$ & 8.0 & --- & 1.000 & 1.00 \\
0.80 & 16 & 1.018 & 1.022 & 1.01 \\
0.90 & 36 & 1.090 & 1.078 & 1.04 \\
0.95 & 76 & 1.196 & 1.189 & 1.09 \\
\bottomrule
\end{tabular}
\end{table}

The crossover $\alpha=1$ occurs at $g=2/3$, marking the boundary of the
Cauchy BTF conjecture's validity range.

\begin{figure}[ht]
\centering
\includegraphics[width=0.75\textwidth]{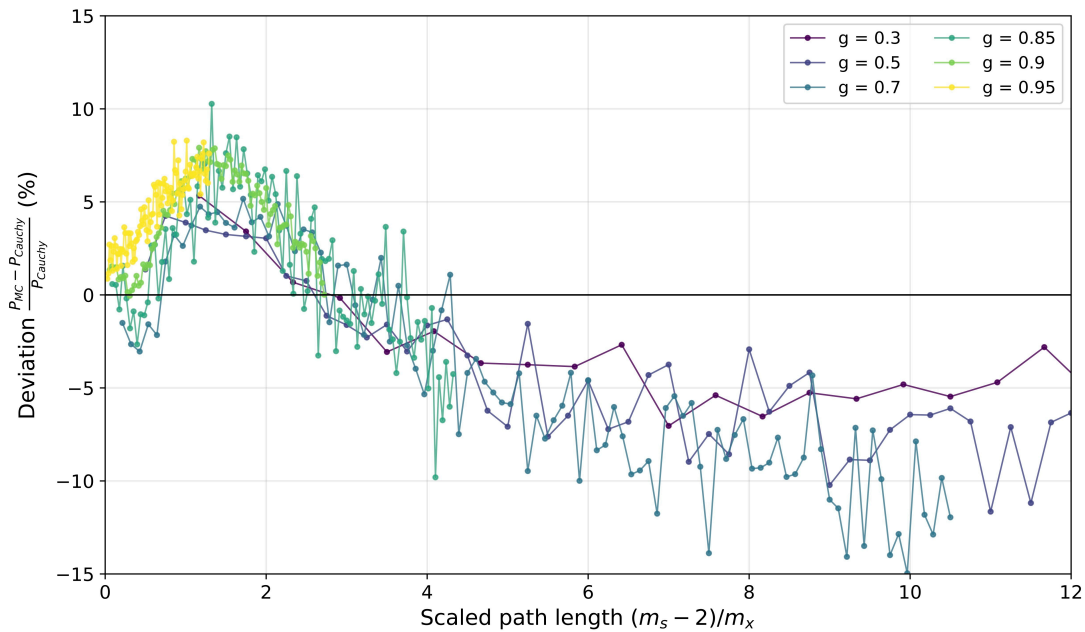}
\caption{Deviations from the Cauchy kernel versus scaled path length. At
high~$g$, Monte Carlo results exceed the Cauchy fit near the peak and fall
below it in the tail---the signature of lighter-than-Cauchy tails.}
\label{fig:deviations}
\end{figure}

The formula contains the factor $1/(1-g)=\ell^*/\ell$, where $\ell$ is the
scattering mean free path and $\ell^*=\ell/(1-g)$ is the transport mean free
path. This ratio appears throughout radiative transfer theory as the natural
length scale for direction randomization~\cite{Chandrasekhar1960}. Its
emergence in the shape parameter suggests a connection to Chandrasekhar's
similarity principle: transport properties depend on $g$ primarily through
$\ell^*$, not $\ell$ and $g$ separately.

We stress that equation~\eqref{eq:alpha} is an empirical fit, not a
derivation. It minimizes squared error across all $n$ and captures the trend;
the coefficient 0.033 is a fitted constant without theoretical explanation.

\begin{figure}[ht]
\centering
\includegraphics[width=\textwidth]{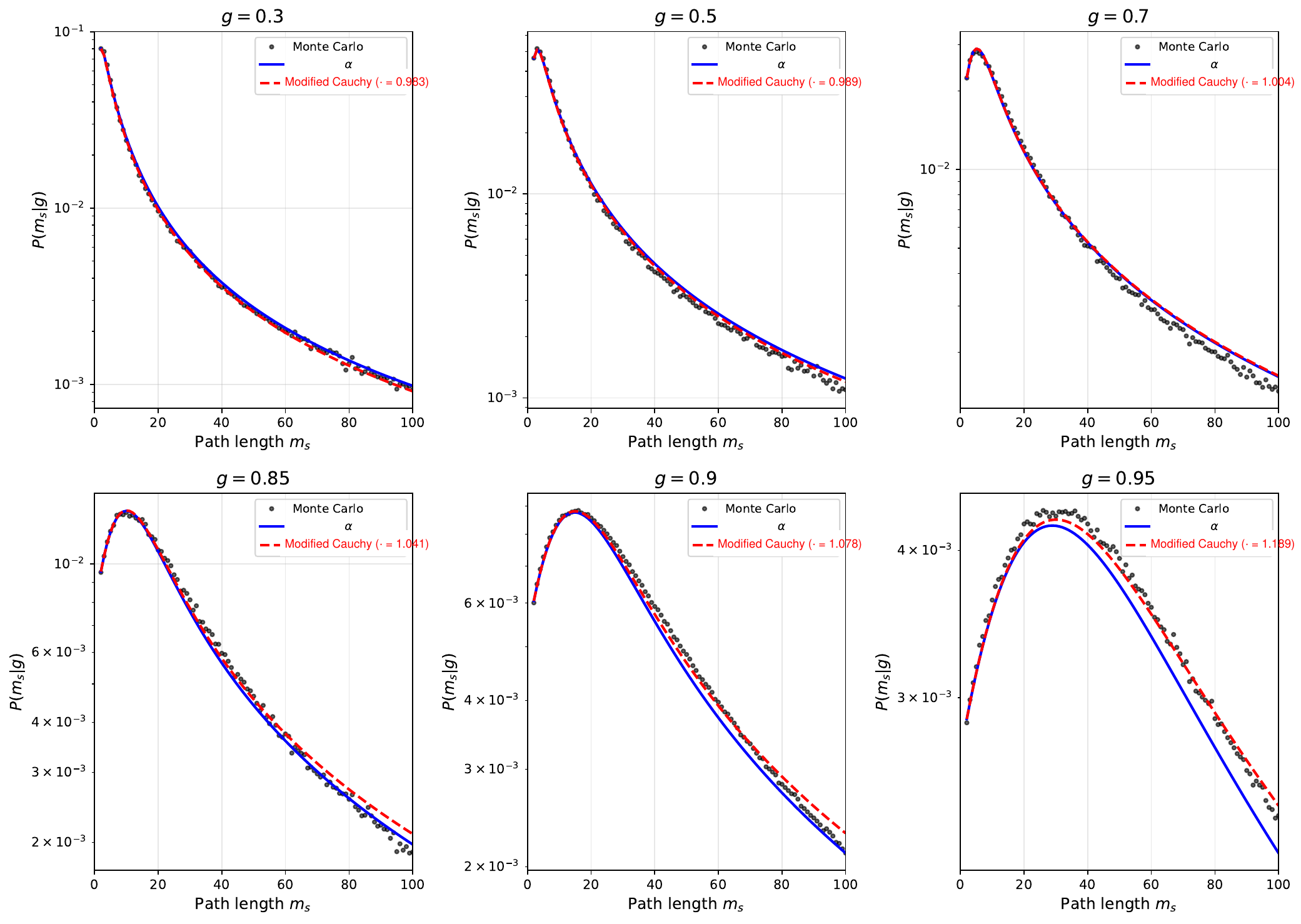}
\caption{Path length distributions $P(m_s|g)$ for six anisotropy values.
Gray points: Monte Carlo. Blue solid: Cauchy kernel ($\alpha=1$). Red dashed:
Modified Cauchy kernel. At $g\le 0.7$, the curves are indistinguishable; at
$g\ge 0.85$, the modified kernel captures the lighter tails.}
\label{fig:pathlen}
\end{figure}

\begin{figure}[ht]
\centering
\includegraphics[width=0.65\textwidth]{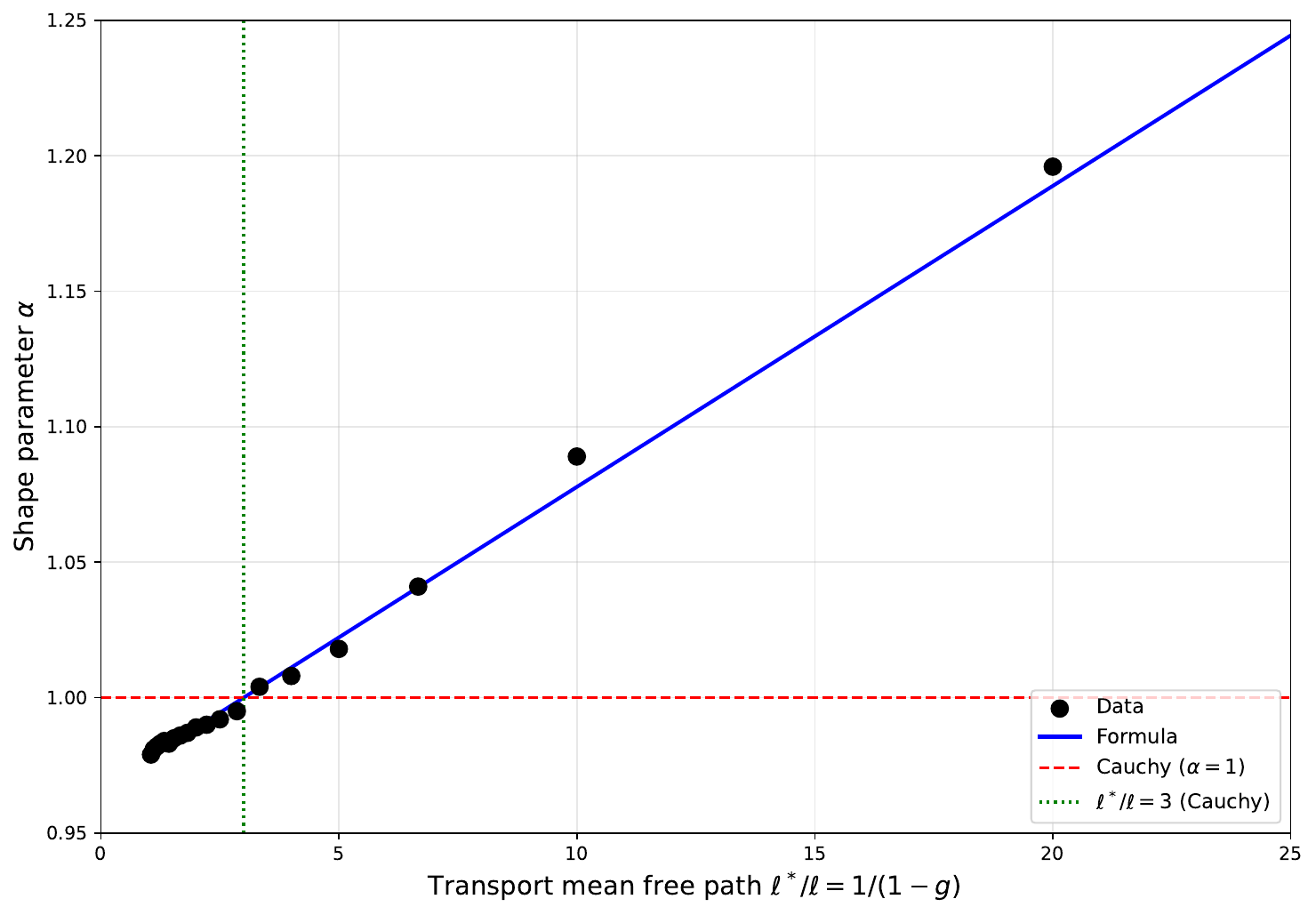}
\caption{Shape parameter $\alpha$ versus transport mean free path ratio
$\ell^*/\ell=1/(1-g)$. The Cauchy case $\alpha=1$ occurs at $\ell^*/\ell=3$.}
\label{fig:alpha}
\end{figure}

\subsection{Practical recommendations}
\label{sec:highg-recommendations}

Guidelines for implementation:
\begin{itemize}
\item $g<0.85$: use the standard Cauchy BTF. Errors remain under 2\%, and the
formula is simpler.
\item $0.85\le g\le 0.95$: use the modified form with $\alpha$ from
equation~\eqref{eq:alpha}. RMSE improves by $\sim$45\%.
\item $g>0.95$: the formula has not been validated beyond $g=0.95$. Monte
Carlo calibration is recommended.
\end{itemize}
For tissue optics ($g\approx 0.9$--$0.95$), the modified kernel extends the
analytical BTF into the regime of primary interest.

\section{Conclusions}
\label{sec:conclusions}

The central empirical result is that the Boundary Truncation Factor for 3D
Henyey--Greenstein scattering takes a Cauchy kernel form, with parameters
$n_0=2$, $m_x=4g/(1-g)$, and $A(g)=1-g(1+g)/2$. Monte Carlo simulations
with $10^{12}$ scattering events confirm this to 1--2\% accuracy for $g<2/3$.
The parsimony is remarkable: just five parameters capture the entire parameter
space ($n=2$--100, $g=0$--0.95).

The BTF corrects for transverse diffusion, which reduces return probability at
large~$n$. The Cauchy form emerges robustly from Monte Carlo. Why this form,
and not another, remains open; proving or refuting the Cauchy BTF conjecture
(Section~\ref{sec:Cauchy}) is the central open problem of this work.

For high anisotropy ($g>2/3$), a modified Cauchy kernel with shape parameter
$\alpha(g)$ extends the theory to $g\approx 0.95$, covering biological tissue.
The high-$g$ extension is phenomenological but constrained by Monte Carlo
data---not a free fit.

The extension to oblique incidence (Section~\ref{sec:oblique}) shows that the
BTF framework is not limited to normal incidence. The key insight is that
$A(g)$ and $m_x(g)$ are angle-independent: only the single-scattering return
probability changes, while the Motzkin counting machinery carries through.
This means the framework handles the full physical problem---reflectance of a
pencil beam at arbitrary incidence angle on a scattering half-space.

Taken together, these results provide an analytical solution of the scalar
radiative transfer equation within this canonical geometry: semi-infinite
half-space, Henyey--Greenstein phase function, collimated beam at arbitrary
incidence, and pure scattering. Because the scattering-order-resolved probability
$P(n,g,\mu_{\mathrm{inc}})$ acts as a generating function---reflectance at
any single-scattering albedo~$a$ follows from
$R=\sum_{n}P(n)\,a^n$---a single evaluation of the framework covers all
albedos. In practice, this replaces $10^8$-trajectory Monte Carlo simulations
with 1D polynomial evaluation, making the approach particularly attractive
for inverse problems requiring thousands of transport
evaluations~\cite{Jacques2013,Modric2009}. The geometry is restricted to
semi-infinite media with no spatial or angular resolution of the exit
distribution; extending to finite slabs and resolved exit quantities remains
future work.

\subsection{Open problems}
\label{sec:open}

Several points lack theoretical foundation:

(1) \textbf{Why Cauchy?}\quad The Cauchy BTF conjecture is supported by
extensive Monte Carlo evidence but is not derived from first principles. The
geometric mechanism---transverse filtering---is clear, but why it produces
precisely the Cauchy form (rather than Gaussian, exponential, or other) is
unexplained.

(2) \textbf{Amplitude formula.}\quad The expression
$A(g)=1-g(1+g)/2$ comes from fitting. A derivation from hemisphere geometry
and flux balance would be more satisfying.

(3) \textbf{Width formula.}\quad The factor 4 in $m_x=4g/(1-g)$ may decompose
as $2\times 2$: one factor from
$\langle\ell^2\rangle/\langle\ell\rangle^2=2$ (exponential path lengths), one
from first-passage geometry. The geometric factor awaits rigorous analysis.

(4) \textbf{Shape parameter formula.}\quad The empirical fit
$\alpha(g)=1+0.033(g-2/3)/(1-g)$ works but is unexplained. The coefficient
0.033 is a fitted constant without evident geometric meaning, unlike the
integer coefficients in the Cauchy kernel.

The integer coefficients (2, 4) are not rounded approximations: fitted values
across 10 independent Monte Carlo runs are consistent with exactly 2 and 4
within statistical uncertainty. This, combined with the Cauchy kernel form,
suggests that the Cauchy BTF conjecture may admit a clean derivation. We leave
this as a challenge for future work.

(5) \textbf{Oblique BTF parameters.}\quad Monte Carlo validation at oblique
angles has confirmed the angle-independence of $A(g)$ and $m_x(g)$.
Systematic characterization across the full $(g,\,\mu_{\mathrm{inc}})$
parameter space remains to be done.

\section*{Acknowledgements}

CZ thanks Dr.~Florence Zeller for discussions on Chebyshev polynomials and
Motzkin structures, Professor Arnold Kim for encouragement and guidance, and
Professor Michel Talagrand for perspective on the difficulty of proving the
Cauchy kernel rigorously.

\appendix
\section{Notation summary}
\label{app:notation}

\begin{table}[ht]
\centering
\caption{Principal notation.}
\label{tab:notation}
\small
\begin{tabular}{lll}
\toprule
Symbol & Definition & Notes \\
\midrule
\multicolumn{3}{l}{\textit{Scattering parameters}} \\
$g$ & $\langle\cos\theta\rangle$ & Anisotropy factor, $g\in[0,1)$ \\
$\mu$ & $\cos\theta$ & Direction cosine \\
$\mu_{\mathrm{inc}}$ & $\cos\theta_{\mathrm{inc}}$ & Incidence cosine, $(0,1]$ \\
$P_{\HG}(\mu;g)$ & Eq.~\eqref{eq:PHG} & Henyey--Greenstein phase function \\
$F(\mu;g)$ & Eq.~\eqref{eq:CDF} & HG cumulative distribution function \\
$S$ & scattering coefficient & Units: inverse length \\
$K$ & absorption coefficient & Units: inverse length \\
$a$ & single-scattering albedo & $S/(S+K)$ \\[4pt]
\multicolumn{3}{l}{\textit{Random walk quantities}} \\
$n$ & scattering order & Number of scattering events \\
$m_s$ & path length & $\equiv n$; standard MC terminology \\
$n_p$ & peak count & Direction reversals \\
$\ell$ & step length & Distance between scattering events \\[4pt]
\multicolumn{3}{l}{\textit{Step probabilities}} \\
$r_b$ & backscattering probability & 3D $\to$ 1D mapped parameter \\
$r$ & backward-step probability & 1D model parameter \\[4pt]
\multicolumn{3}{l}{\textit{BTF parameters}} \\
BTF & Boundary Truncation Factor & Eq.~\eqref{eq:BTF-Cauchy} \\
$A(g)$ & amplitude & $1-g(1+g)/2$ \\
$m_x(g)$ & width parameter & $4g/(1-g)$ \\
$n_0$ & peak location & $=2$ \\
$\alpha(g)$ & shape parameter & $1+0.033(g-2/3)/(1-g)$; Eq.~\eqref{eq:alpha} \\[4pt]
\multicolumn{3}{l}{\textit{Algorithm quantities}} \\
$\mu_b(g)$ & angular threshold & Eq.~\eqref{eq:mub} \\
$p_{r2}(g)$ & two-step return probability & Eq.~\eqref{eq:pr2} \\
$p_{r2}(g,\mu_{\mathrm{inc}})$ & oblique return probability & Eq.~\eqref{eq:pr2-oblique} \\
$I_l(\mu_{\mathrm{inc}})$ & modal geometric integral & Eq.~\eqref{eq:Il} \\[4pt]
\multicolumn{3}{l}{\textit{Combinatorial objects}} \\
$C_n$ & Catalan number & $(2n)!/(n!(n+1)!)$ \\
$M_n(t)$ & Motzkin polynomial & Eq.~\eqref{eq:Motzkin-poly} \\
$T(n,k)$ & Motzkin triangle coefficient & $n!/((n-2k)!\,k!\,(k+1)!)$ \\
\bottomrule
\end{tabular}
\end{table}

\section{First few cases of $P_{\mathrm{1D}}^{\mathrm{marg}}$}
\label{app:cases}

For reference, the first-return probability~\eqref{eq:P1Dmarg} evaluated at
small scattering orders:

\begin{center}
\begin{tabular}{ccl}
\toprule
$m_s$ & $n_p$ range & $P_{\mathrm{1D}}^{\mathrm{marg}}(m_s,r)$ \\
\midrule
2 & 1 & $r/2$ \\
3 & 1 & $r(1-r)/2$ \\
4 & 1--2 & $r(1-r)^2/2 + r^3/8$ \\
5 & 1--2 & $r(1-r)^3/2 + 3r^3(1-r)/8$ \\
6 & 1--3 & $r(1-r)^4/2 + 3r^3(1-r)^2/4 + r^5/16$ \\
\bottomrule
\end{tabular}
\end{center}

\noindent
Verification: $P(2,g)=r_b/2=p_{r2}(g)$ exactly (self-consistency with Step~1).

\section*{AI Disclosure}

An AI-assisted language model (Claude, Anthropic, 2025--2026) was used during
manuscript preparation to assist in checking and verifying portions of the
authors' own mathematical derivations and to improve clarity of presentation.
The AI tool did not generate original results, proofs, or derivations. All
mathematics, analyses, and conclusions were independently derived, reviewed,
and validated by the authors, who take full responsibility for the content of
the manuscript.


\end{document}